\documentclass[%
,secnumarabic%
,amssymb,aps,prl,showpacs,showkeys,nobibnotes]{revtex4}
\usepackage{docs}
\usepackage{bm}%
\usepackage{graphicx}
\expandafter\ifx\csname package@font\endcsname\relax\else
 \expandafter\expandafter
 \expandafter\usepackage
 \expandafter\expandafter
 \expandafter{\csname package@font\endcsname}%
\fi
\DeclareRobustCommand\substyle{\name@idx{document substyle}}%
\DeclareRobustCommand\classoption{\name@idx{document class option}}%
\DeclareRobustCommand\classname{\name@idx{document class}}%
\def\name@idx#1#2{%
 {\ttfamily#2}%
 \index{#2\space#1=\string\ttt{#2}\space#1}\index{#1>#2=\string\ttt{#2}}%
}%
\begin{document}
\title{Spatiotemporal pattern formation of Beddington-DeAngelis-type
predator-prey model}%
\author{Weiming Wang}%
\email{weimingwang2003@163.com} \affiliation{Institute of Nonlinear
Analysis, College of Mathematics and Information Science, Wenzhou
University, Wenzhou, Zhejiang, 325035} \affiliation{Department of
Mathematics, North University of China, Taiyuan, Shan'xi 030051,
P.R. China}
\author{Lei Zhang}
\affiliation{Institute of Nonlinear Analysis, College of Mathematics
and Information Science, Wenzhou University, Wenzhou, Zhejiang,
325035} \affiliation{Department of Mathematics, North University of
China, Taiyuan, Shan'xi 030051, P.R. China}
\author{Yakui Xue}
\affiliation{Department of Mathematics, North University of China,
Taiyuan, Shan'xi 030051, P.R. China}
\author{Zhen Jin}
\affiliation{Department of Mathematics, North University of China,
Taiyuan, Shan'xi 030051, P.R. China}
\date{\today}%

\begin{abstract}
In this paper, we investigate the emergence of a predator-prey model
with Beddington-DeAngelis-type functional response and
reaction-diffusion. We derive the conditions for Hopf and Turing
bifurcation on the spatial domain. Based on the stability and
bifurcation analysis, we give the spatial pattern formation via
numerical simulation, i.e., the evolution process of the model near
the coexistence equilibrium point. We find that for the model we
consider, pure Turing instability gives birth to the spotted
pattern, pure Hopf instability gives birth to the spiral wave
pattern, and both Hopf and Turing instability give birth to
stripe-like pattern. Our results show that reaction-diffusion model
is an appropriate tool for investigating fundamental mechanism of
complex spatiotemporal dynamics. It will be useful for studying the
dynamic complexity of ecosystems.
\end{abstract}



\maketitle \tableofcontents

\section{Introduction}\label{sec1}

Mathematical models have played an important role throughout the
history of ecology. Models in ecology serve a variety of purposes,
which range from illustrating an idea to parameterizing a complex
real-world situation. They are used to make general predictions, to
guide management practices, and to provide a basis for the
development of statistical tools and testable
hypotheses~\citep{ClaudiaNeuhauser,SimonA.Levin01171997,Murray2003}.

A fundamental goal of theoretical ecology is to understand how the
interactions of individual organisms with each other and with the
environment determine the distribution of populations and the
structure of communities~\citep{CANTRELL2003}. As we know, our
ecological environment is a huge and highly complex system. This
complexity arises in part from the diversity of biological species,
and also from the complexity of every individual
organism~\citep{Jost1998,Jost1999}. The dynamic behavior of
predator-prey model has long been and will continue to be one of the
dominant themes in both ecology and mathematical ecology due to its
universal existence and importance~\citep{Berryman,kuang98global}.

Before the 1970s, ecological population models typically used
ordinary differential equations, seeking equilibria and analyzing
stability. The early models provided important insights, such as
when species can stably coexist and when predator and prey density
oscillate over time~\citep{ClaudiaNeuhauser}.

We live in a spatial world, and spatial patterns are ubiquitous in
nature, which modify the temporal dynamics and stability properties
of population density at a range of spatial scales, whose effects
must be incorporated in temporal ecological models that do not
represent space explicitly. And the spatial component of ecological
interactions has been identified as an important factor in how
ecological communities are shaped~\citep{ClaudiaNeuhauser}.
Empirical evidence suggests that the spatial scale and structure of
the environment can influence population interactions and the
composition of communities~\citep{CANTRELL2003}. In recent decades,
the role of spatial effects in maintaining biodiversity has received
a great deal of attention in the literature on
conservation~\citep{Levin1992,medvinsky:311,Murray2003,
Martin2006,Maionchi,liu:031110, Wang,Griffith}.

The past investigations have revealed that spatial inhomogeneities
like the inhomogeneous distribution of nutrients as well as
interactions on spatial scales like migration can have an important
impact on the dynamics of ecological
populations~\citep{medvinsky:311, Murray2003}. In particular it has
been shown that spatial inhomogeneities promote the persistence of
ecological populations, play an important role in speciation and
stabilize population levels~\citep{Martin2006}. Spatial ecology
today is still dominated by theoretical investigations, and
empirical studies that explore the role of space are becoming more
common due to technological advances that allow the recording of
exact spatial locations~\citep{ClaudiaNeuhauser}.

In 1952, Alan Turing, one of the key scientists of 20th century,
mathematically showed that a system of coupled reaction-diffusion
equations could give rise to spatial concentration patterns of a
fixed characteristic length from an arbitrary initial configuration
due to diffusion-driven instability~\citep{Turing1952}. The work by
Turing belongs to the field of pattern formation, a subfield of
mathematical biology. Pattern formation in nonlinear complex systems
is one of the central problems of the natural, social, and
technological sciences. The occurrence of multiple steady states and
transitions from one to another after critical fluctuations, the
phenomena of excitability, oscillations, waves and the emergence of
macroscopic order from microscopic interactions in various nonlinear
nonequilibrium systems in nature and society have been the subject
of many theoretical and experimental studies~\citep{medvinsky:311}.
It has been qualitatively shown that Turing models can indeed
imitate biological patterns~\citep{Teemu2004}.

The study of biological pattern formation has gained popularity
since the 1970s, and Segel and Jackson~\citep{Segel1972} were the
first to apply Turing's ideas to a problem in population dynamics:
the dissipative instability in the predator-prey interaction of
phytoplankton and herbivorous copepods with higher herbivore
motility. At the same time, Gierer and Meinhardt~\citep{Gierer1972}
gave a biologically justified formulation of a Turing model and
studied its properties by employing numerical simulations. Levin and
Segel~\citep{Levin1976} suggested this scenario of spatial pattern
formation was a possible origin of planktonic patchiness.

Spatial patterns and aggregated population distributions are common
in nature and in a variety of spatiotemporal models with local
ecological interactions~\citep{Pascual}. And the understanding of
patterns and mechanism of species spatial dispersal is an issue of
current interest in conservation biology and
ecology~\citep{Levin1992,medvinsky:311,Andrew2006}. It arises from
many ecological applications, and in particular, plays a major role
in connection to biological invasion and epidemic spread and so
on~\citep{Andrew2006,Levin1992,medvinsky:311,Maini2004,Maini2006,Martin2006,Murray2003,ClaudiaNeuhauser,liu:031110,Garvie}.
The field of research on pattern formation modeled by
reaction-diffusion systems, which provides a general theoretical
framework for describing pattern formation in systems from many
diverse disciplines including biology~\citep{Levin1992,
Klausmeier06111999, medvinsky:311,
PhysRevE.74.011914,Hardenberg,CANTRELL2003, PhysRevE.66.010901,
Pascual1993, Pascual, Wang,Shuji}, chemistry\citep{Ouyang1991,
Vanag2000, yang:7259,Karen, PhysRevLett.88.208303,
yang:178303,SanderJDE,yang:026211, yang2006},
physics\citep{PhysRevE.64.026219,Arecchi,
RevModPhys.65.851,PhysRevE.70.066202},
epidemiology~\citep{liu:031110,LiuJSM} and so on, seems to be a new
increasingly interesting area, particularly during the last decade.

In general, a classical predator-prey model can be written as the
form~\citep{Arditi1989,David2002}:
$$
\begin{array}{l}
 \dot{N}=Nf(N)-Pg(N,P),\qquad
 \dot{P}=h[g(N,P),P] P.
\end{array}
$$
where $N$ and $P$ are prey and predator densities, respectively,
$f(N)$ the prey growth rate, $g(N, P)$ the functional response,
e.g., the prey consumption rate by an average single predator, and
$h[g(N, P), P]$ the per capita growth rate of predators (also known
as the ``predator numerical response"), which obviously increases
with the prey consumption rate. The most widely accepted assumption
for the numerical response is the linear one~\citep{David2002}:
$$
h[g(N, P), P]=\varepsilon g(N, P)-\eta
$$
where $\eta$ is a per capita predator death rate and $\varepsilon$
the conversion efficiency of food into offspring.

In population dynamics, a functional response $g(N,P)$ of the
predator to the prey density refers to the change in the density of
prey attached per unit time per predator as the prey density
changes~\citep{Abram2000, Ruan:1445}. There have been several famous
functional response types: Holling types I---III~\citep{Holling1,
Holling2}; Hassell-Varley type~\citep{Hassell}; Beddington-DeAngelis
type by Beddington~\citep{Beddington} and DeAngelis \emph{et
al}~\citep{DeAngelis} independently; the Crowley-Martin
type~\citep{Crowley}; and the recent well-known ratio-dependence
type by Arditi and Ginzburg~\citep{Arditi1989} later studied by
Kuang and Beretta~\citep{kuang98global}. Of them, the Holling type
I-III is labeled ``prey-dependent" and other types that consider the
interference among predators are labeled
``predator-dependent"~\citep{Arditi1989}.

In our previous work~\citep{Wang}, we studied the spatiotemporal
complexity of a ratio-dependent predator-prey model with
Michaelis-Menten-type functional response. Compare
Michaelis-Menten-type functional response $$g(N,P)=\frac{\alpha
N}{P+\alpha hN}\quad$$ ({$\alpha, h$ are positive constants,
$\alpha$ capture rate and $h$ handling time}) with
Beddington-DeAngelis-type functional response
\begin{equation}\label{eq:0}
g(N,P)={\frac {\beta N}{B+N+wP}}\\[4pt]
\end{equation}
($\beta, B$ are positive constants, $\beta$ a maximum consumption
rate, $B$ a saturation constant, $w$ a predator interference
parameter, a constant. $w<0$ is the case where predators benefit
from cofeeding). Some scholars~\citep{Skalski,David2002,Liu2006}
indicat that where there is a small difference between the
denominators, there is a world between them from the biological
point.

If predators do not waste time interacting with one another or if
their attacks are always successful and instantaneous, i.e., $w=0$
in (~\ref{eq:0}), then a Holling type II functional response is
obtained:
$$g(N)=\frac{\beta N}{B+N}.$$

If $B=0$ in (\ref{eq:0}), the Michaelis-Menten-type functional
response is obtained.

Furthermore, in~\citep{Skalski}, by comparing the statistical
evidence from nineteen categories of predator-prey systems with
three predator-dependent functional responses, Skalski and Gilliam
pointed out that the predator-dependent functional responses could
provide better descriptions of predator feeding over a range of
predator-prey abundance, and in some cases, the
Beddington-DeAngelis-type functional response performed even
better~\citep{Liu2006}.

Some progress has been seen in the study of predator-prey model with
Beddington-DeAngelis-type functional
response~\citep{Jost1998,Huisman,Hwang,Kar2007,
Cantrell2001,David2002,Fan2004,Dobromir2005,Liu2006,Cui2006,Wang2006,Wang2007}.
However, the research on considering reaction-diffusion to such a
model, to our knowledge, seems rare.

In this paper, we report a study of Turing pattern formation in a
two-species reaction-diffusion predator-prey model with
Beddington-DeAngelis-type functional response. In the next section
we give a brief stability and bifurcation analysis of the model.
Then, we present and discuss the results of numerical simulations,
which is followed by the last section, i.e., conclusions and
remarks.

\section{Stability and bifurcation analysis}

In this paper, we mainly focus on the following predator-prey model
with Beddington-DeAngelis-type functional response and
reaction-diffusion:
\begin{eqnarray}\label{eq:1}
 \begin{array}{l}
 \frac{\partial N}{\partial t}=r \left( 1-{\frac {N}{K}} \right) N-{\frac {\beta
 N}{B+N+wP}}P+d_1\nabla^2N,\qquad
 \frac{\partial P}{\partial t}={\frac {\varepsilon\beta N}{B+N+wP}}P-\eta P+d_2\nabla^2P.\\[8pt]
 \end{array}
\end{eqnarray}
where $t$ denotes time and $N, P$ stand for prey and predator
density, respectively. All parameters are positive constants, $r$
standing for maximum per capita growth rate of the prey, $\beta$
capture rate, $\eta$ predator death rate, $w$ a predator
interference parameter and $K$ carrying capacity, which is the
nonzero equilibrium population size. The diffusion coefficients are
denoted by $d_1$ and $d_2$, respectively.
$\nabla^2=\frac{\partial}{\partial x^2}+\frac{\partial}{\partial
y^2}$ is the usual Laplacian operator in two-dimensional space.

The first step in analyzing the model is to determine the equilibria
(stationary states) of the non-spatial model obtained by setting
space derivatives equal to zero, i.e.,
\begin{eqnarray}\label{eq:100}
 \begin{array}{l}
 r \left( 1-{\frac {N}{K}} \right) N-{\frac {\beta
 N}{B+N+wP}}P=0,\qquad
 {\frac {\varepsilon\beta N}{B+N+wP}}P-\eta P=0.
 \end{array}
\end{eqnarray}

In fact, physically, an equilibrium represents a situation without
``life". It may mean no motion of a pendulum, no reaction in a
reactor, no nerve activity, no flutter of an airfoil, no laser
operation, or no circadian rhythms of biological
clocks~\citep{Seydel}. And at each equilibrium point, the movement
of the population dynamics vanishes.

Eqs.(\ref{eq:100}) has at most three equilibria, which correspond to
spatially homogeneous equilibria of the full model, Eq.(\ref{eq:1}),
in the positive quadrant:

(i) $(0, 0)$ (total extinct) is a saddle point.

(ii) $(K, 0)$ (extinct of the predator, or prey-only) is a stable
node if $\varepsilon\,\beta<\eta$  or  if $\varepsilon\,\beta>\eta$
and $K<{\frac {\eta\,B}{\varepsilon\,\beta-\eta}}$; a saddle if
$\varepsilon\,\beta<\eta$ and $K>{\frac
{\eta\,B}{\varepsilon\,\beta-\eta}}$; a saddle-node if
$\varepsilon\,\beta<\eta$ and $K={\frac
{\eta\,B}{\varepsilon\,\beta-\eta}}$.

(iii) a nontrivial stationary state $(N^*, P^*)$ (coexistence of
prey and predator), where
$$
\begin{array}{l}
N^*=\frac{1}{2rw\varepsilon}\Bigl(K(rw\varepsilon-\varepsilon\beta+\eta)+\sqrt{{K}^{2}(rw\varepsilon-\varepsilon\,\beta+\eta)^{2}+4\,rKw\varepsilon\eta
B}\Bigr),\\[8pt]
P^*=\frac{(\beta\varepsilon-\eta)}{w\eta}N^*-\frac{B}{w}.
\end{array}
$$

To perform a linear stability analysis, we linearize the dynamic
system \ref{eq:1} around the equilibrium point $(N^*, P^*)$ for
small space- and time-dependent fluctuations and expand them in
Fourier space
$$
N(\vec{x},t)\sim N^*e^{\lambda t}e^{i\vec{k}\cdot\vec{x}},\quad
P(\vec{x},t)\sim P^*e^{\lambda t}e^{i\vec{k}\cdot\vec{x}},
$$
and obtain the characteristic equation
\begin{equation}\label{eq:8}
|A-k^2D-\lambda I|=0,
\end{equation}
where $D=\text{diag}(d_1, d_2)$ and the Jacobian matrix $A$ is given
by
$$
A=\left(\begin
{array}{cc}\partial_nf&\partial_pf\\\noalign{\medskip}\partial_ng&\partial_pg\end
{array} \right)_{(N^*, P^*)}=\left(\begin
{array}{cc}f_n&f_p\\\noalign{\medskip}g_n&g_p\end {array} \right).
$$

Now Eq.(\ref{eq:8}) can be solved, yielding the so called
characteristic polynomial of the original problem (\ref{eq:1}):
\begin{equation}\label{eq:11}
\lambda^2-\text{tr}_k\lambda+\Delta_k=0,
\end{equation}
where $$\begin{array}{l}
\text{tr}_k=f_n+g_p-k^2(d_1+d_2)=\text{tr}_0-k^2(d_1+d_2),\\[8pt]
\Delta_k=f_ng_p-f_pg_n-k^2(f_nd_2+g_pd_1)+k^4d_1d_2=\Delta_0-k^2(f_nd_2+g_pd_1)+k^4d_1d_2.
\end{array}
$$

The roots of Eq.(\ref{eq:11}) yield the dispersion relation
\begin{equation}\label{eq:13}
\lambda(k)=\frac{1}{2}\Bigl(\text{tr}_k\pm\sqrt{\text{tr}_k^2-4\Delta_k}\,\,
\Bigr).
\end{equation}

We know that one type of instability (or bifurcation) will break one
type of symmetry of a system, i.e., in the bifurcation point, two
equilibrium states intersect and exchange their stability.
Biologically speaking, this bifurcation corresponds to a smooth
transition between equilibrium states. The Hopf bifurcation is
space-independent and it breaks the temporal symmetry of a system
and gives rise to oscillations that are uniform in space and
periodic in time. The Turing bifurcation breaks spatial symmetry,
leading to the formation of patterns that are stationary in time and
oscillatory in space.

The Hopf instability or bifurcation is an important instability in
reaction-diffusion systems for which the conditions result in a
stable limit cycle (oscillations). In terms of the linearized
problem (Eq.\ref{eq:8}) the onset of Hopf instability corresponds to
the case, when a pair of imaginary eigenvalues cross the real axis
from the negative to the positive side. The Hopf bifurcation of a
equilibrium state is reflected by a transition between stationary
and periodic behavior. If the system is in a stable equilibrium
before the bifurcation, the stability is lost at the bifurcation
point. As a result the system abundance of species start to
oscillate periodically. And this situation occurs only when the
diffusion vanishes. Mathematically speaking, the Hopf bifurcation
occurs when $ \text{Im}(\lambda(k))\neq 0, \quad
\text{Re}(\lambda(k))=0\,\,\text{at}\,\,\,k=0. $. Then we can get
the critical value of the transition, Hopf bifurcation parameter---
$K$, equals
\begin{eqnarray}\label{eq:15}
K_H={\frac {B \left( w\varepsilon\,\eta-\beta\,\varepsilon-\eta
\right) ^{2}}{
 \left( w\varepsilon-1 \right)  \left( {\eta}^{2}-{\beta}^{2}{\varepsilon}^{
2}+rw{\varepsilon}^{2}\beta+{\varepsilon}^{2}\beta\,w\eta-w\varepsilon\,{\eta}^
{2} \right) }}
\end{eqnarray}

At the Hopf bifurcation threshold, the temporal symmetry of the
system is broken and gives rise to uniform oscillations in space and
periodic oscillations in time with the frequency
$$
\omega_H=\text{Im}(\lambda(k))=\sqrt{\Delta_0},
$$
where
$$\begin{array}{l}\Delta_0=-\frac{\Bigl(K(\eta-\beta\,\varepsilon)( K
(rw\varepsilon-\beta\,\varepsilon+\eta)^{2}+\eta\,\delta-\beta\,\delta\,\varepsilon+4\,r
w\varepsilon\,\eta\,B+r\delta\,\varepsilon\,w
)+2\,r\delta\,\eta\,B\varepsilon\,w \Bigr)\eta}{\beta\,K
\varepsilon^{2}w( rKw\varepsilon-\beta\,K\varepsilon+K\eta+\delta)
},
\end{array}
$$
and
$$\begin{array}{l}\delta=\Bigl({r}^{2}{K}^{2}{w}^{2}{\varepsilon}^{2}-2\,r{K}^{2}w{\varepsilon}^{2}
\beta+2\,r{K}^{2}w\varepsilon\,\eta+{\beta}^{2}{K}^{2}{\varepsilon}^{2}-2\,
\beta\,{K}^{2}\varepsilon\,\eta+{K}^{2}{\eta}^{2}+4\,rw\varepsilon\,K\eta\,B
\Bigr)^{1/2}.\end{array}$$
The corresponding wavelength is
$$
\lambda_H=\frac{2\pi}{\omega_H}=\frac{2\pi}{\sqrt{\Delta_0}}\,.
$$

The Turing instability is not dependent upon the geometry of the
system but only upon the reaction rates and diffusion. And it cannot
be expected when the diffusion term is absent and it can occur only
when the activator (e.g., $N$) diffuses more slowly than the
inhibitor (e.g., $P$). Linear analysis above shows that the
necessary conditions for yielding Turing patterns are given by
$$
\begin{array}{l}f_n+g_p<0,\quad
f_ng_p-f_pg_n>0,\quad d_2g_p+d_1f_n>0,\quad
(d_2f_n+d_1g_p)^2>4d_1d_2(f_ng_p-f_pg_n). \end{array}
$$
Mathematically speaking, as $d_1\ll d_2$, the Turing bifurcation
occurs when
$$
\text{Im}(\lambda(k))=0, \quad
\text{Re}(\lambda(k))=0\,\,\,\text{at}\,\,\,k=k_T\neq 0,
$$
and the wavenumber $k_T$ satisfies
$k_T^2=\sqrt{\frac{\Delta_0}{d_1d_2}}.$
At the Turing bifurcation
threshold, the spatial symmetry of the system is broken and the
patterns are stationary in time and oscillatory in space with the
corresponding wavelength
\begin{equation}\label{eq:19}
\lambda_T=\frac{2\pi}{k_T}.
\end{equation}
And the critical value of Turing bifurcation parameter $K$ takes the
following form:
\begin{equation}\label{eq:20}
\begin{array}{l}
K_T=\frac{F_{{1}}A+F_{{2}}}{{G_{{1}}}A+{G_{2}}},
\end{array}\end{equation} where
\begin{widetext}
$${
\begin{array}{l}
A=\left(-\eta d_1\,(\varepsilon\beta-\eta)(-
\varepsilon\,d_2\beta-\eta\,d_2+\varepsilon w\eta d_1
\right)^2\left(-\varepsilon\eta\,d_1\,rw-\varepsilon
\eta\,d_1\,\beta+\varepsilon\,r^2
d_2\,w-\varepsilon\,rd_2\,\beta+d_1\,\eta^2-rd_2\,\eta)
\right)^{1/2},
\\[4pt]
F_{1}=-\left(\eta^2( d_1\varepsilon\,w-d_2)^2+
\varepsilon\,d_2\beta(\varepsilon d_2\beta+2\eta d_2 +6\varepsilon
w\eta d_1)\right)^{2}((-2{\varepsilon}^{2}\beta\,wd_2(
2\,w\varepsilon\,{\eta}^{2}d_1d_2-4\,w{\varepsilon}^{2}\eta\,d_1d_2\beta-{\eta}^{2}{d_1}^{
2}{\varepsilon}^{2}{w}^{2}\\[4pt]
\qquad\,\,
+{\beta}^{2}{d_{{2}}}^{2}{\varepsilon}^{2}-{d_2}^{2}{\eta}^{2})r)+2\,\varepsilon\,d_{{2}}\beta\,(\varepsilon
\,\beta-\eta)({\varepsilon}^{2}( d_{{2}}\beta+w\eta \,d_{{1}})
^{2}+\eta\,d_{{2}}(\eta\,d_{{2}}-2\,\varepsilon
\,w\eta\,d_{{1}}+2\,\varepsilon\,d_{{2}}\beta)))rB
,\\[4pt]
F_{2}=-({\eta}^{2}(d_{{1}}\varepsilon\,w-d_{{2}})^{2}+\varepsilon\,d_{{2}}\beta\,(
\varepsilon\,d_{{2}}\beta+2\,\eta\,d_{{2}}+6\,\varepsilon\,w\eta\,d_{{1}}
))^{2}(\eta\,\beta \,{\varepsilon}^{2}wd_{{2}}(
\varepsilon\,d_{{2}}\beta+\eta\,d_{{2}}- \varepsilon\,w\eta\,d_{{1}}
)({\varepsilon}^{3}d_{{1}}w( 3\,d_{{2}}\beta+w\eta\,d_{{1}}) ^{2}\\[4pt]
\qquad\,\, +{d_{{2}}}^{3} (
\eta+\varepsilon\,\beta)^{2}-\varepsilon\,d_{{1}}{\eta}^{2}w d_{{2}}
( d_{{1}}\varepsilon\,w+d_{{2}})){r}^{2}- \eta\,(
\varepsilon\,\beta-\eta)( \varepsilon\,d_{{2}}
\beta+\eta\,d_{{2}}-\varepsilon\,w\eta\,d_{{1}})( {\varepsilon
}^{4}{\eta}^{3}{w}^{4}{d_{{1}}}^{4}+{\varepsilon}^{3}{\eta}^{2}{w}^{3}d_{
{2}}( -4\,\eta\\[4pt]
\qquad\,\, +9\,\varepsilon\,\beta) {d_{{1}}}^{3}+3\,{
\varepsilon}^{2}\eta\,{w}^{2}{d_{{2}}}^{2}( -5\,\eta\,\varepsilon\,
\beta+2\,{\eta}^{2}+9\,{\varepsilon}^{2}{\beta}^{2}) {d_{{1}}}^{2}
+\varepsilon\,w{d_{{2}}}^{3}(11\,\varepsilon\,\beta-4\,\eta) (
\eta+\varepsilon\,\beta) ^{2}d_{{1}}+{d_{{2}}}^{4}(
\eta+\varepsilon\,\beta) ^{3} ) r\\[4pt]
\qquad\,\, +2\,\eta\,\beta\,d_{{1}} \varepsilon\,d_{{2}}(
\varepsilon\,\beta-\eta)^{2}(
\varepsilon\,d_{{2}}\beta+\eta\,d_{{2}}-\varepsilon\,w\eta\,d_{{1}}
) ( {\varepsilon}^{2}( d_{{2}}\beta+w\eta\,d_{{1}}) ^{2}
+\eta\,d_{{2}}( \eta\,d_{{2}}-2\,\varepsilon\,w\eta\,d_{{1}}+2\,
\varepsilon\,d_{{2}}\beta))) rB
,\\[4pt]
G_1=G_{{11}}G_{{12}},\\[4pt]
G_{11}=4\,\varepsilon\,d_{{2}}\beta\,(-(
{\varepsilon}^{2}\beta\,d_{{2 }}( d_{{2}}\beta-4\,w\eta\,d_{{1}} )
-{\eta}^{2}( d_ {{1}}\varepsilon\,w-d_{{2}}) ^{2} )
w\varepsilon\,r+( \varepsilon\,\beta-\eta)({\varepsilon}^{2}(
d_{{2}} \beta+w\eta\,d_{{1}}) ^{2}+\eta\,d_{{2}}( \eta\,d_{{2}}-2
\,\varepsilon\,w\eta\,d_{{1}}\\[4pt]
\qquad\,\, +2\,\varepsilon\,d_{{2}}\beta))
),\\[4pt]
G_{12}=({\varepsilon}^{3}d_{{1}}w(
3\,d_{{2}}\beta+w\eta\,d_{{1}})^{2}+{d_{{2}}}^{3}(
\eta+\varepsilon\,\beta)^{2}-
\varepsilon\,d_{{1}}{\eta}^{2}wd_{{2}}(
d_{{1}}\varepsilon\,w+d_{{2}}
))d_{{2}}w\beta\,{\varepsilon}^{2}{r}^{2}-(\varepsilon\,\beta-\eta)
({\varepsilon}^{3}{d_{{2}}}^{3} ( d_{{2}}+11\,d_{{1}}\varepsilon\,w
){\beta}^{3}\\[4pt]
\qquad\,\,+3\,{d_{{2}} }^{2}\eta\,{\varepsilon}^{2}(
3\,d_{{1}}\varepsilon\,w+d_{{2}}
)^{2}{\beta}^{2}+3\,\varepsilon\,{\eta}^{2}d_{{2}}(3\,d_{{
1}}\varepsilon\,w+d_{{2}})(d_{{1}}\varepsilon\,w-d_{{2}})
^{2}\beta+{\eta}^{3}(d_{{1}}\varepsilon\,w-d_{{2}}
)^{4})r+2\,\varepsilon\,\beta\,d_{{1}}d_{{2}}(
\varepsilon\,\beta-\eta)^{2}\\[4pt]\qquad\,\,
\left( {\varepsilon}^{2}( d_{{2}} \beta+w\eta\,d_{{1}})
^{2}+\eta\,d_{{2}}( \eta\,d_{{2}}-2
\,\varepsilon\,w\eta\,d_{{1}}+2\,\varepsilon\,d_{{2}}\beta) \right)
,\\[4pt]
G_{2}=G_{{21}}G_{{22}},\\[4pt]
G_{21}=\varepsilon\,d_{{2}}\beta+\eta\,d_{{2}}-\varepsilon\,w\eta\,d_{{1}},\\[4pt]
G_{22}=g_{{0}}+g_{{1}}r+g_{{2}}{r}^{2}+g_{{3}}{r}^{3}+g_{{4}}{r}^{4},\\[4pt]
g_{0}=8\,{\varepsilon}^{2}{\beta}^{2}\eta\,{d_{{1}}}^{2}{d_{{2}}}^{2}
(\varepsilon\,\beta-\eta)^{4}( {\varepsilon}^{2}(d_{{2}}
\beta+w\eta\,d_{{1}}) ^{2}-2\,\varepsilon\,d_{{1}}{\eta}^{2}wd_{{2
}}+{d_{{2}}}^{2}{\eta}^{2}+2\,\beta\,\varepsilon\,\eta\,{d_{{2}}}^{2}
)^{2},\\[4pt]
g_{1}=4\,\varepsilon\,\beta\,d_{{1}}d_{{2}}(\varepsilon\,\beta-\eta
)^{3}({\varepsilon}^{2}(d_{{2}}\beta+w\eta\,d_{{1}})
^{2}-2\,\varepsilon\,d_{{1}}{\eta}^{2}wd_{{2}}+{d_{{2}}}^{2}{
\eta}^{2}+2\,\beta\,\varepsilon\,\eta\,{d_{{2}}}^{2})( - (
d_{{1}}\varepsilon\,w-d_{{2}})^{4}{\eta}^{4}-2\,\epsilon
\,\beta\,d_{{2}}( 3\,d_{{1}}\varepsilon\,w+d_{{2}})\\[4pt]
\qquad\,\,( d_{{1}}\varepsilon\,w-d_{{2}}) ^{2}{\eta}^{3}
-16\,{\varepsilon}^{3}{ \beta}^{2}d_{{1}}w{d_{{2}}}^{2}(
d_{{1}}\varepsilon\,w+d_{{2}})
{\eta}^{2}-2\,{\varepsilon}^{3}{d_{{2}}}^{3}{\beta}^{3}( -
d_{{2}}+5\,d_{{1}}\varepsilon\,w )
\eta+{d_{{2}}}^{4}{\varepsilon}^{4} {\beta}^{4}),\\[4pt]
\end{array}}$$
\end{widetext}
\begin{widetext}
$${
\begin{array}{l}
g_{2}=(\varepsilon\,\beta-\eta)^{2}((d_{{1}} \varepsilon\,w-d_{{2}}
)^{8}{\eta}^{7}+6\,\varepsilon\,\beta\,d_{{2}}(
3\,d_{{1}}\varepsilon\,w+d_{{2}})(d_{{1}}\varepsilon
\,w-d_{{2}})^{6}{\eta}^{6}+{\beta}^{2}{d_{{2}}}^{2}{\varepsilon}^{2}(
151\,{\varepsilon}^{2}{w}^{2}{d_{{1}}}^{2}+106\,\varepsilon\,d_{{
1}}wd_{{2}}+15\,{d_{{2}}}^{2})\\[4pt]
\qquad\,\,(d_{{1}}\varepsilon\,w-d_{{2
}})^{4}{\eta}^{5}+4\,{\varepsilon}^{3}{d_{{2}}}^{3}{\beta}^{3}(
156\,{\varepsilon}^{3}{d_{{1}}}^{3}{w}^{3}+133\,{\varepsilon}^{2}{w}
^{2}{d_{{1}}}^{2}d_{{2}}+58\,wd_{{1}}\varepsilon\,{d_{{2}}}^{2}+5\,{d_{{2
}}}^{3})(d_{{1}}\varepsilon\,w-d_{{2}}) ^{2}{\eta}^
{4}+{d_{{2}}}^{4}{\varepsilon}^{4}{\beta}^{4}(
3\,d_{{1}}\varepsilon\, w\\[4pt]
\qquad\,\,+d_{{2}})( 389\,{\varepsilon}^{3}{d_{{1}}}^{3}{w}^{3}+117
\,{\varepsilon}^{2}{w}^{2}{d_{{1}}}^{2}d_{{2}}+183\,wd_{{1}}\varepsilon\,{d_
{{2}}}^{2}+15\,{d_{{2}}}^{3}) {\eta}^{3}+2\,{\varepsilon}^{5}{
\beta}^{5}{d_{{2}}}^{5}(357\,{\varepsilon}^{3}{d_{{1}}}^{3}{w}^{3}
+297\,{\varepsilon}^{2}{w}^{2}{d_{{1}}}^{2}d_{{2}}\\[4pt]
\qquad\,\,+47\,wd_{{1}}\varepsilon\, {d_{{2}}}^{2}+3\,{d_{{2}}}^{3})
{\eta}^{2}+{\varepsilon}^{6}{\beta }^{6}{d_{{2}}}^{6}(
153\,{\varepsilon}^{2}{w}^{2}{d_{{1}}}^{2}+{d_{
{2}}}^{2}-10\,\varepsilon\,d_{{1}}wd_{{2}})
\eta-12\,{\varepsilon}^{8 }{\beta}^{7}wd_{{1}}{d_{{2}}}^{7}),
\\[4pt]
g_{3}=-2\,{\varepsilon}^{2}\beta\,wd_{{2}}(\varepsilon\,\beta-\eta )
(( d_{{1}}\varepsilon\,w+d_{{2}})(d_{{1}}\varepsilon\,w-d_{{2}})
^{6}{\eta}^{6}+\varepsilon\,\beta\,d_{{2}}(13\,{\varepsilon}^{2}{w}^{2}{d_{{1}}}^{2}
+6\,\varepsilon\,d_{{1}}wd_{{2}}+5\,{d_{{2}}}^{2})(
d_{{1}}\varepsilon\,w-d_{{2}})^{4}{\eta}^{5}\\[4pt]
\qquad\,\,+2\,{\beta}^{2}{d_{{2}}}^{2}{\varepsilon}^{2}(38\,{\varepsilon}^{3}{d_{{1}}}^{3}{w}^{3}+{\varepsilon}^{2}{w}^{2}{d
_{{1}}}^{2}d_{{2}}+20\,wd_{{1}}\varepsilon\,{d_{{2}}}^{2}+5\,{d_{{2}}}^{3
})( d_{{1}}\varepsilon\,w-d_{{2}}) ^{2}{\eta}^{4}+2
\,{\varepsilon}^{3}{\beta}^{3}{d_{{2}}}^{3}(5\,{d_{{2}}}^{4}+117\,
{\varepsilon}^{4}{w}^{4}{d_{{1}}}^{4}
\\[4pt]\qquad\,\,
+44\,\varepsilon\,d_{{1}}w{d_{{2}}}^{3}+14\,{d_{{2}}}^{2}{d_{{1}}}^{2}{w}^{2}{\varepsilon}^{2}-52\,d_{{2}}{d_{{1
}}}^{3}{w}^{3}{\varepsilon}^{3})
{\eta}^{3}+{d_{{2}}}^{4}{\varepsilon }^{4}{\beta}^{4}(
227\,{\varepsilon}^{2}{w}^{2}{d_{{1}}}^{2}d_{{2}}
+329\,{\varepsilon}^{3}{d_{{1}}}^{3}{w}^{3}+79\,wd_{{1}}\varepsilon\,{d_{{2}
}}^{2}\\[4pt]
\qquad\,\,+5\,{d_{{2}}}^{3})
{\eta}^{2}+{\varepsilon}^{5}{\beta}^{5}{d
_{{2}}}^{5}(14\,\varepsilon\,d_{{1}}wd_{{2}}+{d_{{2}}}^{2}+121\,{
\varepsilon}^{2}{w}^{2}{d_{{1}}}^{2})
\eta-6\,{\varepsilon}^{7}d_{{1} }w{d_{{2}}}^{6}{\beta}^{6}) ,
\\[4pt]
g_{4}={\varepsilon}^{4}{\beta}^{2}{w}^{2}{d_{{2}}}^{2}((
{\varepsilon
}^{2}{w}^{2}{d_{{1}}}^{2}+6\,\varepsilon\,d_{{1}}wd_{{2}}+{d_{{2}}}^{2}
)(wd_{{1}}\varepsilon-d_{{2}})^{4}{\eta}^{5}+4\,
\varepsilon\,\beta\,d_{{2}}(2\,{\varepsilon}^{3}{d_{{1}}}^{3}{w}^{3}+
13\,{\varepsilon}^{2}{w}^{2}{d_{{1}}}^{2}d_{{2}}+{d_{{2}}}^{3} )
(wd_{{1}}\varepsilon-d_{{2}})^{2}{\eta}^{4}\\[4pt]
\qquad\,\,+2\,{\beta}^{2}
{d_{{2}}}^{2}{\varepsilon}^{2}(3\,{d_{{2}}}^{4}+52\,d_{{2}}{d_{{1}
}}^{3}{w}^{3}{\varepsilon}^{3}-6\,{d_{{2}}}^{2}{d_{{1}}}^{2}{w}^{2}{
\varepsilon}^{2}+4\,\varepsilon\,d_{{1}}w{d_{{2}}}^{3}+11\,{\varepsilon}^{4}{w}
^{4}{d_{{1}}}^{4}){\eta}^{3}+4\,{\varepsilon}^{3}{\beta}^{3}{d_{{
2}}}^{3}(-9\,{\varepsilon}^{2}{w}^{2}{d_{{1}}}^{2}d_{{2}}+{d_{{2}}
}^{3}\\[4pt]
\qquad\,\,+13\,{\varepsilon}^{3}{d_{{1}}}^{3}{w}^{3}+11\,wd_{{1}}\varepsilon\,{d_
{{2}}}^{2}){\eta}^{2}+{d_{{2}}}^{4}{\varepsilon}^{4}{\beta}^{4}(
113\,{\varepsilon}^{2}{w}^{2}{d_{{1}}}^{2}+{d_{{2}}}^{2}+22\,
\varepsilon\,d_{{1}}wd_{{2}})
\eta-4\,{d_{{2}}}^{5}{\varepsilon}^{6}w d_{{1}}{\beta}^{5}),
\end{array}}
$$
\end{widetext}

\begin{figure*}[htp]
\includegraphics[width=15cm,height=9cm]{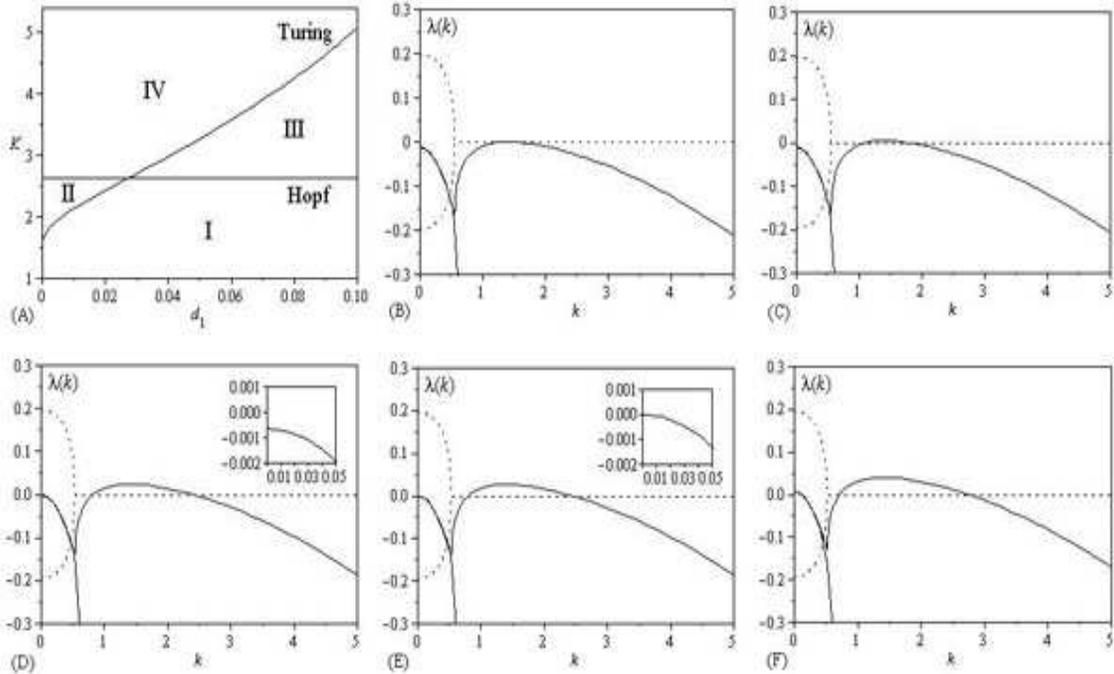}
\caption{\label{fig1} (A) $K-d_1$ Bifurcation diagram for the model
\ref{eq:1} with $r=0.5$, $\varepsilon=1$, $\beta=0.6$, $B=0.4$,
$\eta=0.25$, $w=0.4$ and $d_2=1$. Hopf and Turing bifurcation lines
separate the parameter space into four domains. And the Hopf-Turing
bifurcation point is $(d_1, K)=(0.02742, 2.63158)$. In (B)---(F),
$\text{Re}(\lambda(k))$ and $\text{Im}(\lambda(k))$ are shown by
solid curves and dotted curves, respectively. The other parameters
are: $d_1=0.01$, the bifurcation parameter $K$: (B) 2.131712170, the
critical value of $K_T$); (C) 2.2; (D) 2.6; (E) 2.631578947 the
critical value of $K_H$); (F) 3.0.}
\end{figure*}

In the following, linear stability analysis yields the bifurcation
diagram with $r=0.5$, $\varepsilon=1$, $\beta=0.6$, $B=0.4$,
$\eta=0.25$, $w=0.4$,$d_2=1$ shown in Fig.~\ref{fig1}(A).

The Hopf bifurcation line and the Turing bifurcation curve separate
the parametric space into four distinct domains. In domain I,
located below all two bifurcation lines, the steady state is the
only stable solution of the system. Domain II is the region of pure
Turing instability, while domain III is the region of pure Hopf
instability. In domain IV, which is located above all two
bifurcation lines, both Hopf and Turing instability occur.

To see the relation between the real and the imaginary parts of the
eigenvalue $\lambda(k)$, we plot in Fig.~\ref{fig1}(B)--(F) the real
and the imaginary parts of the eigenvalue at different $K$ with
$r=0.5$, $\varepsilon=1$, $\beta=0.6$, $B=0.4$, $\eta=0.25$, $w=0.4$
$d_1=0.01$ and $d_2=1$ for the system \ref{eq:1}.

From the definition of Hopf and Turing bifurcation, we know that the
relation between the real, the imaginary parts of the eigenvalue
$\lambda(k)$ determine the bifurcation type. The relation between
$\text{Re}(\lambda(k))$, $\text{Im}(\lambda(k))$ and $k$ are shown
in figure~\ref{fig1}(B)--(F). Figure~\ref{fig1}(B) illustrate the
case of parameter locate in domain I in figure~\ref{fig1}(A),
$K=2.131712170$, the critical value of Turing bifurcation, in this
case, $\text{Re}(\lambda(k))>0$ at $k=0$ while
$\text{Im}(\lambda(k))\neq0$. In figure~\ref{fig1}(C)(D), $K=2.2$
and $K=2.6$, the parameter locate in domain II, the pure Turing
instability occurs, one can see that at $k=0$,
$\text{Re}(\lambda(k))=0$, $\text{Im}(\lambda(k))\neq0$.
Figure~\ref{fig1}(E), $K=2.631578947$, the critical value of Hopf
bifurcation, in this case, $\text{Re}(\lambda(k))=0$ at $k=0$ while
$\text{Im}(\lambda(k))\neq0$. When $K=3.0$, parameter locate in
domain IV, figure~\ref{fig1}(F) indicate that at $k=0$,
$\text{Re}(\lambda(k))>0$, $\text{Im}(\lambda(k))\neq0$.

\section{Spatiotemporal pattern formation}

In this section, we perform extensive numerical simulations of the
spatially extended model (\ref{eq:1}) in two-dimensional spaces, and
the qualitative results are shown here. All our numerical
simulations employ the periodic Neumann (zero-flux) boundary
conditions with a system size of $200\times200$ space units and
$r=0.5$, $\varepsilon=1$, $\beta=0.6$, $B=0.4$, $\eta=0.25$, $w=0.4$
$d_1=0.01$ and $d_2=1$. The Eq.\ref{eq:1} are solved numerically in
two-dimensional space using a finite difference approximation for
the spatial derivatives and an explicit Euler method for the time
integration with a time stepsize of $\Delta t=0.01$ and space
stepsize (lattice constant) $\Delta h=0.25$ (see, for details,
~\citep{Garvie}). The scale of the space and time are average to the
Euler method. The initial density distribution corresponds to random
perturbations around the stationary state $(N^*, P^*)$ in model
(\ref{eq:1}) with a variance significantly lower than the amplitude
of the final patterns, which seems to be more general from the
biological point of view. When the system reached a stable state
(stationary or oscillatory), we took a snapshot with yellow levels
linearly proportional to the free species density and red
corresponding to high while blue corresponding to low.

In the numerical simulations, different types of dynamics are
observed and we have found that the distributions of predator and
prey are always of the same type. Consequently, we can restrict our
analysis of pattern formation to one distribution. In this section,
we show the distribution of prey, for instance.

\begin{figure*}[htp]
\includegraphics[width=13cm]{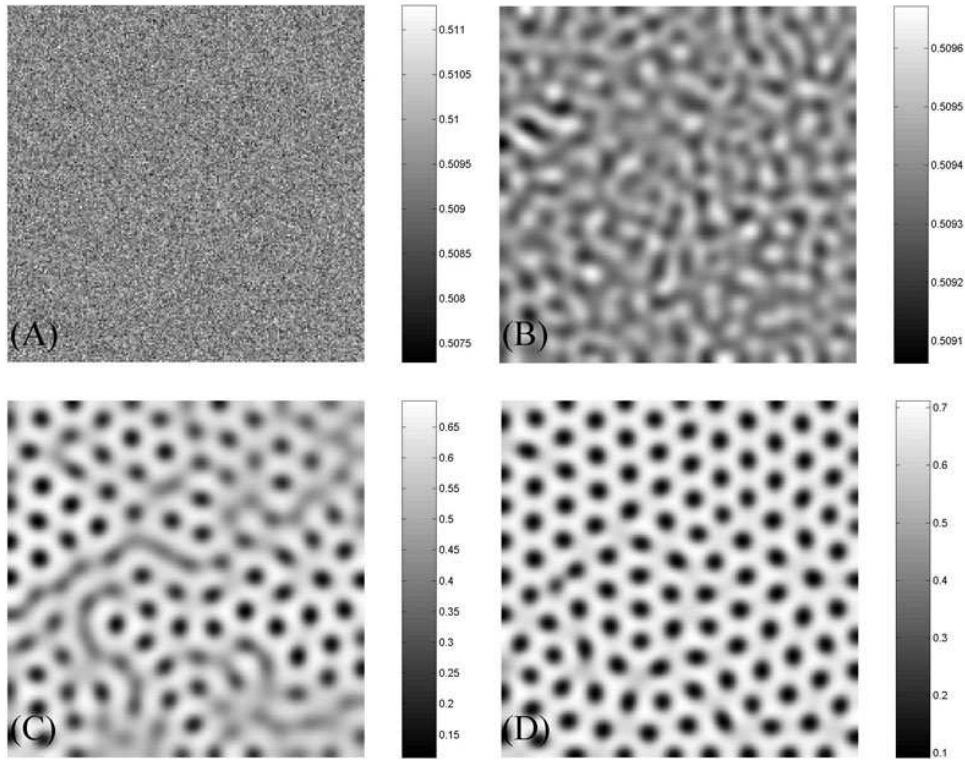}
\caption{\label{fig2} Dynamics of the time evolution of the prey of
model \ref{eq:1} with $d_1=0.01,\,K_T<K=2.2<K_H$. (A) 0 iteration,
(B) 10000 iterations, (C) 200000 iterations, (D) 300000 iterations.}
\end{figure*}

\begin{figure*}[htp]
\includegraphics[width=12.5cm]{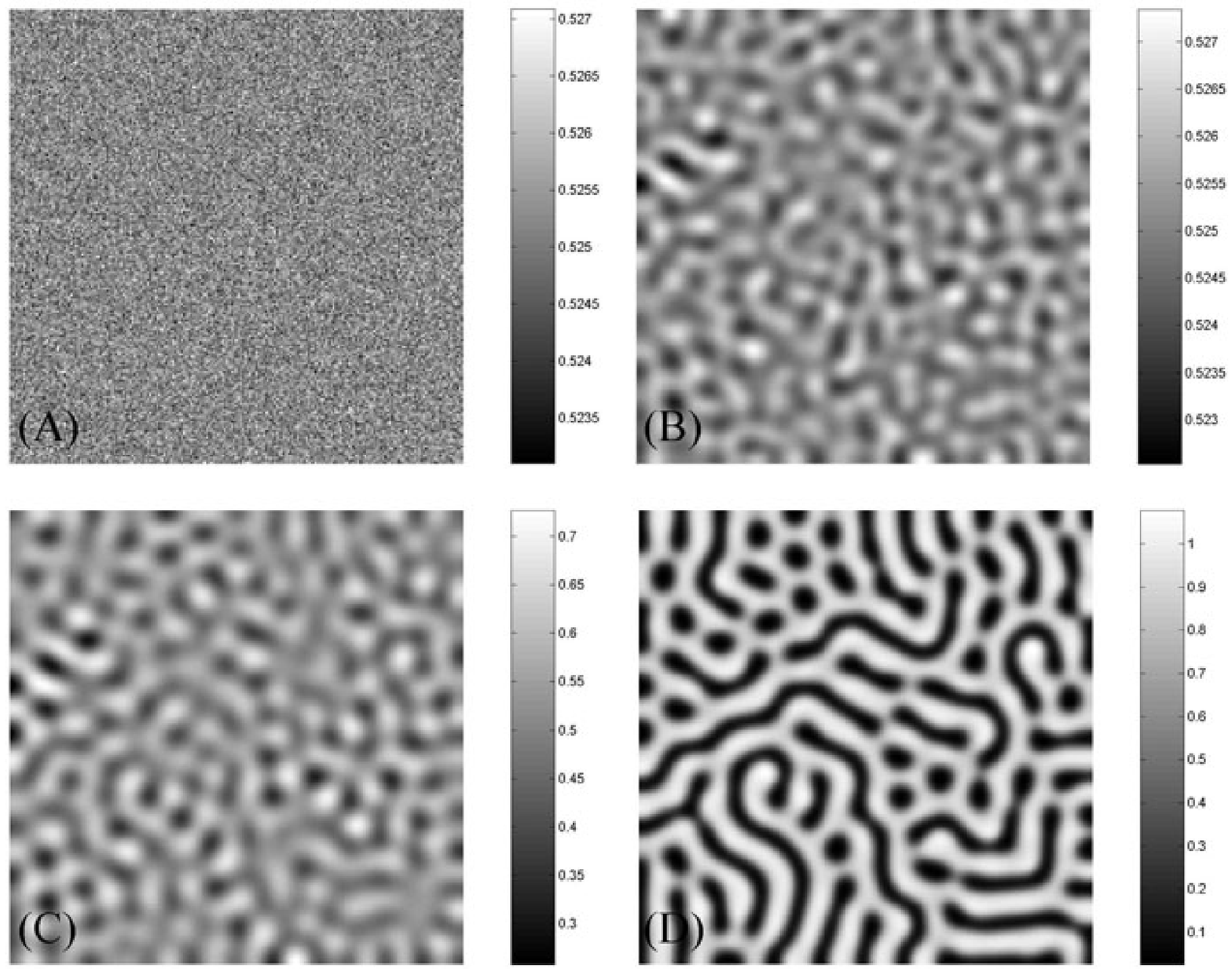}
\caption{\label{fig3} Dynamics of the time evolution of the prey of
model \ref{eq:1} with $d_1=0.01,\,K_T<K=2.6<K_H$. (A) 0 iteration,
(B) 10000 iterations, (C) 30000 iterations, (D) 100000 iterations.}
\end{figure*}

\begin{figure*}[htp]
\includegraphics[width=12.5cm]{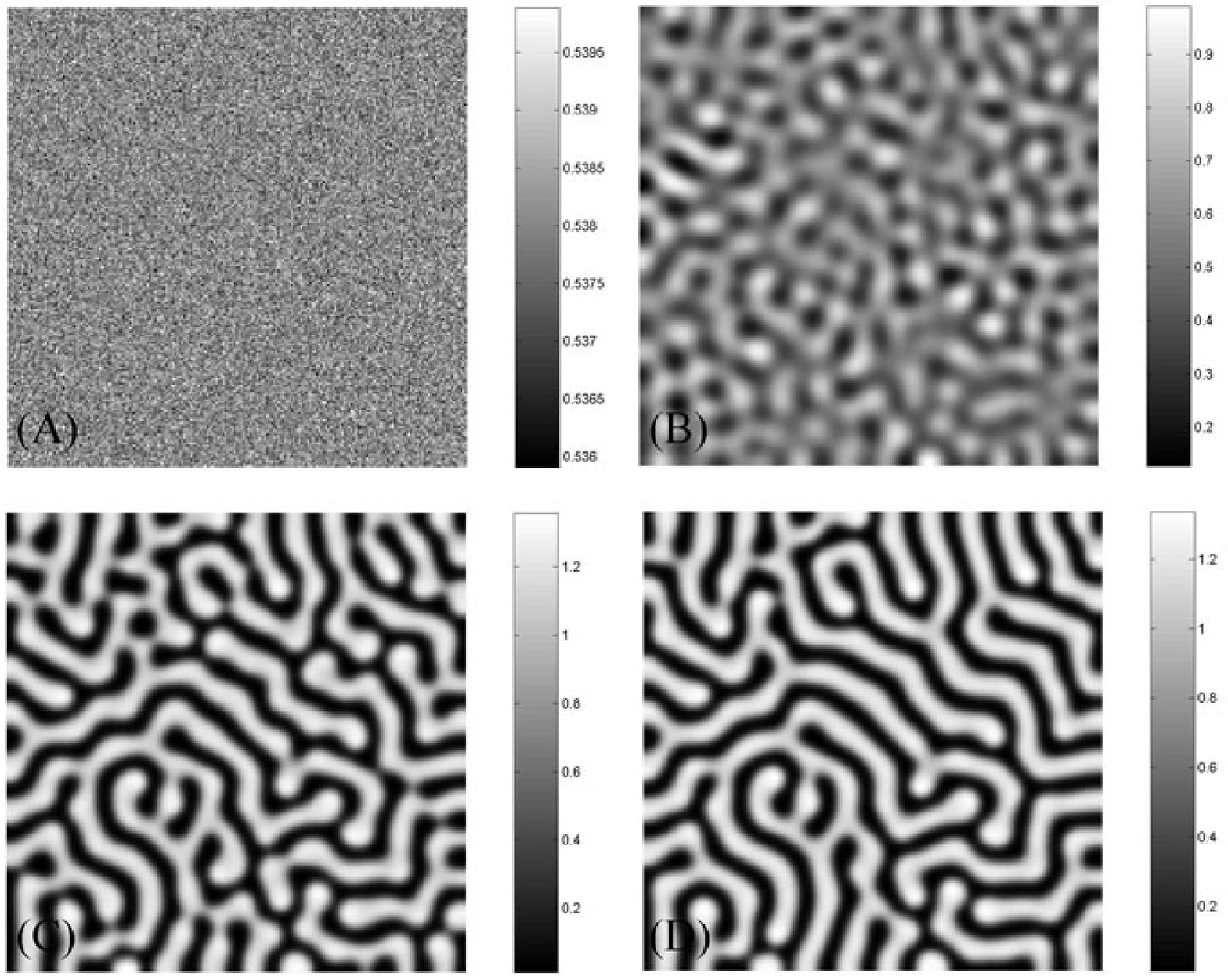}
\caption{\label{fig4} Dynamics of the time evolution of the prey of
model \ref{eq:1} with $d_1=0.01,\,K_T<K_H<K=3.0$. (A) 0 iteration,
(B) 20000 iterations, (C) 50000 iterations, (D) 100000 iterations.}
\end{figure*}

\begin{figure*}[htp]
\includegraphics[width=12cm]{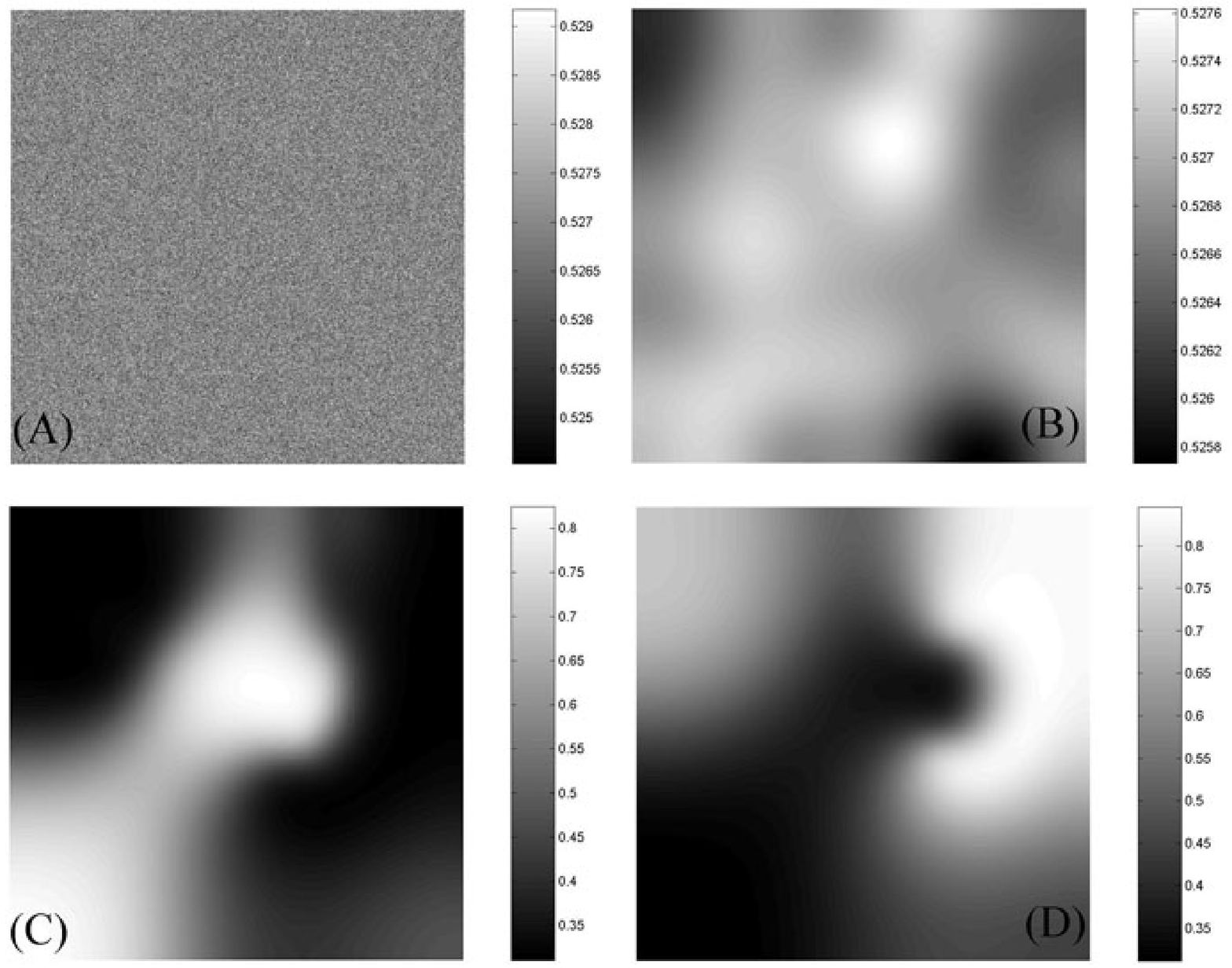}
\caption{\label{fig5} Dynamics of the time evolution of the prey of
model \ref{eq:1} with $d_1=0.04,\,K_T<K_H<K=2.65$. (A) 0 iteration,
(B) 30000 iterations, (C) 70000 iterations, (D) 100000 iterations.}
\end{figure*}

\begin{figure*}[htp]
\includegraphics[width=17cm]{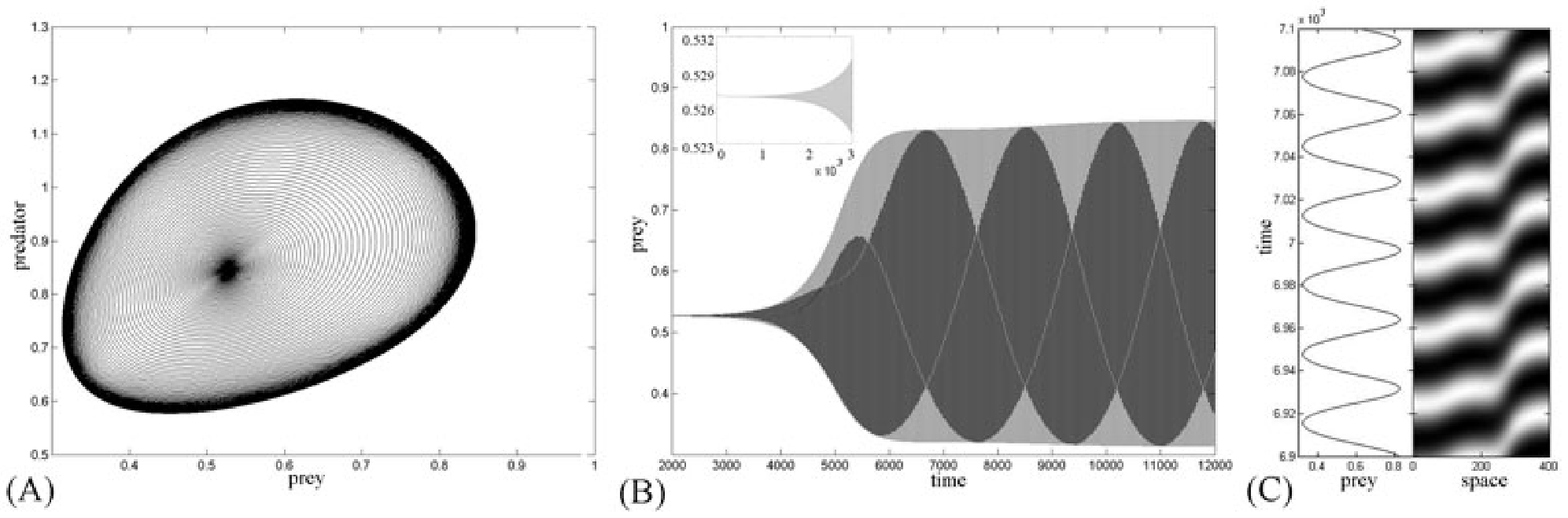}
\caption{\label{fig6} Dynamical behaviors of model \ref{eq:1}. (A)
Phase portrait; (B) Time-series plot; (C) Space-time plots
corresponding to Fig.~\ref{fig5}(C). The parameters are the same as
those in Fig~\ref{fig5}.}
\end{figure*}

Fig.~\ref{fig2} shows the evolution of the spatial pattern of prey
at 0, 10000, 200000 and 300000 iterations, with random small
perturbation of the stationary solution $(N^*, P^*)=(0.5094,
0.7829)$ of the spatially homogeneous systems when $K=2.2$, located
in domain II, slightly more than the Turing bifurcation threshold
$K_T=2.1317$ and less than the Hopf bifurcation threshold
$K_H=2.6316$. In this case, one can see that for model~\ref{eq:1},
the random initial distribution (c.f., Fig.~\ref{fig2}(A)) leads to
the formation of a regular macroscopic spotted pattern which
prevails over the whole domain at last, and the dynamics of the
system does not undergo any further changes (c.f.,
Fig.~\ref{fig2}(D)).

Fig.~\ref{fig3} shows the evolution of the spatial pattern of prey
at 0, 10000, 30000 and 100000 iterations when $K=2.6$ which is more
than $K_T$ and slightly less than $K_H$. Although the dynamics of
the system starts from the stationary solution $(N^*, P^*)=(0.5252,
0.8382)$, there is an essential difference in the case above. From
the snapshots, one can see that the steady state of spotted pattern
and the stripe-like pattern coexist (c.f., Fig.~\ref{fig3}(D)).

In Fig.~\ref{fig4}, $K_H<K=3.0$, i.e., parameters in domain IV, both
Hopf and Turing instability occur. In this case, $(N^*,
P^*)=(0.5380, 0.8831)$. One can see that the evolution of the
spatial pattern of prey at 0, 20000, 50000 and 100000 iterations.
After the spatial chaos patterns (c.f., Fig.~\ref{fig4}(B)), a
regular stationary stripe-like spatial state emerges (c.f.,
Fig.~\ref{fig4}(D)).

When $K=2.65,\, d_1=0.04$, i.e., parameters in domain III (c.f.,
Fig.~\ref{fig1}(A)), pure Hopf instability occurs. As an example,
the formation of a regular macroscopic two-dimensional spatial
pattern, the spiral pattern, is shown in Fig.~\ref{fig5} with a
system size of $400\times 400$ space units. One can see that for
model~\ref{eq:1}, the random initial distribution around the steady
state $(N^*, P^*)=(0.5270, 0.8443)$, a unstable focus of
model~\ref{eq:1}, leads to the formation of the spiral pattern in
the domain (c.f., Fig.~\ref{fig5}(D)). In other words, in this
situation, spatially uniform steady-state predator-prey coexistence
is no longer. Small random fluctuations will be strongly amplified
by diffusion, leading to nonuniform population distributions. From
the analysis in section II, we find with these parameters in domain
III, the pattern formation, i.e., the spiral pattern, arises from
pure Hopf instability. In order to make it clearer, in
Fig.~\ref{fig6}, we show phase portrait (Fig.~\ref{fig6}(A)), time
series plot (Fig.~\ref{fig6}(B)) and space-time plots
(Fig.~\ref{fig6}(C)), a one-dimensional example corresponding to
Fig.~\ref{fig5}(C). And the method of space-time plots is that let
$y$ be a constant (here, $y=200$, the center line of each
snapshots), from each pattern snapshots, choose the line $y=200$,
and pile these lines in-time-order. The space-time plots show the
evolution process of the prey $N$ throughout time $t$ and space $x$.
Fig.~\ref{fig6}(A) exhibits the ``local" phase plane of the system
obtained in a fixed point $(0.5270, 0.8443)$ inside the region
invaded by the irregular spatiotemporal oscillations, and the
trajectory fills nearly the whole domain inside the limit cycle.
Fig.~\ref{fig6}(B) illustrates the evolution process of prey density
with time, periodic oscillations in time finally. From
Fig.~\ref{fig6}(C), one can clearly see that a spiral wave emerges,
and the system gives rise to uniform oscillations in space and
periodic oscillations in time.

Comparing Fig.~\ref{fig2}, Fig.~\ref{fig3} with Fig.~\ref{fig4},
Fig.~\ref{fig5}, we can see that the bifurcation parameter $K$
determines the type of the pattern formation even with the same
parameters, e.g., $r$, $\varepsilon$, $\beta$, $B$, $\eta$, $w$,
$d_2$. In domain II of Fig.~\ref{fig1}(A), the closer $K$ is to
$K_T$, the more distinct the spotted spatial pattern becomes (c.f.,
Fig.~\ref{fig2}(D)). When $K$ is much closer to $K_H$, the spotted
and stripe-like patterns coexist (c.f., Fig.~\ref{fig3}(D)). When
$K$ is bigger than $K_H$ and far away from the Turing bifurcation
value $K_T$, the distinct stripe-like pattern emerges. When $K$ is
bigger than $K_H$ and smaller than $K_T$, the spiral wave pattern
occurs (c.f., Fig.~\ref{fig5}(C,D)). So we may draw a conclusion
that for model~\ref{eq:1} pure Turing instability gives birth to the
spotted pattern, pure Hopf bifurcation gives birth to the spiral
wave pattern, and both of them give birth to the stripe-like
pattern.

\section{Conclusions and remarks}

In this paper, we have presented a theoretical analysis of
evolutionary processes that involves organisms distribution and
their interaction of spatially distributed population with local
diffusion. And the numerical simulations were consistent with the
predictions drawn from the bifurcation analysis, i.e., Hopf
bifurcation and Turing instability. In the domain II of
Fig.~\ref{fig1}(A), the stationary state of periodic spotted pattern
exists when the parameters are near the Turing bifurcation line,
while near the Hopf bifurcation line, both the spotted pattern and
the stripe-like pattern coexist. When the parameters are located in
domain III, pure Hopf instability occurs, and the spiral wave
pattern emerges. When the parameters are located in domain IV, both
Hopf and Turing instability occur, and the stationary state of
stripe-like pattern exists.

Turing instability is relevant not only in reaction-diffusion
systems, but also in describing other dissipative structures, which
can be understood in terms of diffusion-driven instability. In
addition to the biological relevance of Turing systems, their
ability to generate structure is of great interest from the point of
view of physics. There are various physical systems that show
similar phenomena, although the underlying mechanisms can be very
different. Thus, most of the research in the field relies on
experiments and numerical simulations justified by an analytical
examination. In addition, in simulations one may study pattern
formation under constraints that are beyond the reach of experiments
and the numerical data is also easy to analyze.

In Ref.~\citep{David2002}, it's indicated that the basic idea of
diffusion-driven instability in a reaction-diffusion model can be
understood in terms of an activator-inhibitor system. And a random
increase of activator species (prey, $N$) has a positive effect on
the creation rate of both activator and inhibitor (predator, $P$)
species. In other words, random fluctuations may cause a nonuniform
prey density. This elevated prey density has a positive effect both
on prey and predator population growth rates. Following D. Alonso
\emph{et al}~\citep{David2002}, we give the discussion to
model(~\ref{eq:1}). From Eqs.(~\ref{eq:1}), we can obtain the
following equations:
\begin{eqnarray}\label{eq:30}
\begin{array}{l}
 \frac{1}{N}\frac{\partial N}{\partial t}=r(1-\frac{N}{K})-\frac{\beta P}{B+N+wP},
 \qquad
 \frac{1}{P}\frac{\partial P}{\partial t}=\frac{\varepsilon\beta N}{B+N+wP}-\eta.
\end{array}
\end{eqnarray}
Similar to Ref.~\citep{David2002}, the first equation in
Eqs.(\ref{eq:30}) is a one-humped function of prey density, the
numerical result can be found in~\citep{Huisman}, and prey growth
rate can be increased by a higher local prey density at least in a
range of parameter values. On the other hand, the second equation in
Eqs.(\ref{eq:30}), i.e., predator numerical response, is an
ever-increasing function of $N$, and high prey density always has a
positive influence on predator growth. More importantly, inhibitor
species (predator, $P$) must diffuse faster than activator species
(prey, $N$), for an increment in inhibitor species may have a
negative effect on formation rate of both species. Thus, as random
fluctuations increase local prey density over its equilibrium value,
prey population undergoes an accelerated growth. Simultaneously,
predator population also increases, but as predators diffuse faster
than prey, they disperse away from the center of prey outbreaks. If
relative diffusion ($d_2/d_1$) is large enough, prey growth rate
will reach negative values and prey population will be driven by
predators to a very low level in those regions. The final result is
the formation of patches of high prey density surrounded by areas of
low prey density. Predators follow the same pattern.

In Ref.~\citep{Neuhauser01171997}, Neuhauser and Pacala formulated
the Lotka-Volterra model as a spatial one. They found the striking
result that the coexistence of patterns is actually harder to get in
the spatial model than in the non-spatial one. One reason can be
traced to how local interaction between individual members of the
species are represented in the model. In this study, our results
show that the predator-prey model with Beddington-DeAngelis-type
functional response and reaction-diffusion (e.g., Eqs.~\ref{eq:1})
also represents rich spatial dynamics, such as spotted pattern,
stripe-like pattern, coexistence of both stripe-like and spotted
pattern, spiral pattern, etc. It will be useful for studying the
dynamic complexity of ecosystems or physical systems. In
Ref.~\citep{CANTRELL2003}, it is indicated that reaction-diffusion
models provide a way to translate local assumptions about the
movement, mortality, and reproduction of individuals into global
conclusions about the persistence or extinction of populations and
the coexistence of interacting species. They can be derived
mechanistically via rescaling from models of individual movement
which are based on random walks, i.e., small random perturbation of
the stationary solution $(N^*, P^*)$ of the spatially homogeneous
model (\ref{eq:1}). Reaction-diffusion system, i.e., model
(\ref{eq:1}), is spatially explicit and typically incorporate
quantities such as dispersal rates, local growth rates, and carrying
capacities as parameters which may vary with location or time.

More interesting, when the parameters are located in domain III,
pure Hopf instability occurs, and the spiral wave pattern emerges
(c.f., Fig.~\ref{fig5}, Fig.~\ref{fig6}(C)). To our knowledge, we
haven't got any report about one system that has spotted,
stripe-like and spiral pattern formation meantime. It's well known
that spirals and curves are the most fascinating clusters to emerge
from the predator-prey model. A spiral will form from a wave front
when the rabbit line (which is leading the front) overlaps the
pursuing line of predator. The prey on the extreme end of the line
stop moving as there are no predator in their immediate vicinity.
However the prey and the predator in the center of the line continue
moving forward. This forms a small trail of prey at one (or both)
ends of the front. These prey start breeding and the trailing line
of prey thickens and attracts the attention of predator at the end
of the fox line that turn towards this new source of prey. Thus a
spiral forms with predator on the inside and prey on the outside. If
the original overlap of prey occurs at both ends of the line a
double spiral will form. Spirals can also form as a prey blob
collapses after predator eat into it~\citep{Gurney,Hawick2006}.
Thus, reaction-diffusion system provides a good framework for
studying questions about the ways that habitat geometry and the size
or variation in vital parameters influence population dynamics.

On the other hand, one can see that although there is a small
difference between the denominators of the functional responses of
Michaelis-Menten-type and those of Beddington-DeAngelis-type, there
is an enormous gap between them in the process of computations. The
Turing bifurcation expression of Michaelis-Menten-type predator-prey
system is simple~\citep{Wang}, while from (\ref{eq:20}), one can see
that the Turing bifurcation analysis requires huge-sized
computations, so we have to obtain more help via computers.

In fact, computer aided analysis is useful for nonlinear analysis.
And computers have played an important role throughout the history
of ecology. Today, numerical simulations also play an important role
in spatial ecology. There are some international mathematical
softwares, such as {\tt Matlab, Maple, Mathematica}, etc. We have
finished all our symbolic computations in {\tt Maple} and obtained
our pattern snapshots (i.e., numerical simulations) in {\tt Matlab}
for {\tt Maple} is more superior in symbolic computations while {\tt
Matlab} is more superior in numerical computations. \vskip 0.5cm

\textit{Acknowledgments} This work was supported by the National
Natural Science Foundation of China (10471040) and the Youth Science
Foundation of Shanxi Provence (20041004).


\end{document}